\documentclass[aps,preprint,superscriptaddress,showpacs]{revtex4}
\usepackage{graphicx}
\usepackage{dcolumn}
\usepackage{amsmath}

\begin{document}
\bibliographystyle{apsrev}

\title{Multibranch entrainment and slow evolution 
among branches in coupled oscillators 
}
 
\author{Toru Aonishi}

\affiliation{Brain Science Institute, RIKEN, 
2-1 Hirosawa, Wako-shi, Saitama, 351-0198, Japan 
}
\author{Masato Okada}

\affiliation{ERATO Kawato Dynamic Brain Project, 2-2 Hikaridai, Seika-
cho, 
Soraku-gun, Kyoto 619-0288, Japan 
}

\affiliation{Brain Science Institute, RIKEN, 
2-1 Hirosawa, Wako-shi, Saitama, 351-0198, Japan 
}

\date{\today}

\begin{abstract}
In globally coupled oscillators, it is believed that 
strong higher harmonics of 
coupling functions are essential for {\it multibranch entrainment}
(MBE), in which there exist many stable states, whose number scales as
 $\sim$ $O(\exp N)$ (where $N$ is the system size).
The existence of MBE implies the non-ergodicity of the system.
Then, because this apparent breaking of ergodicity is caused by 
{\it microscopic} energy barriers, this seems to be  
in conflict with a basic principle of statistical physics.
In this paper, using macroscopic dynamical theories, 
we demonstrate that there is no such ergodicity breaking, and 
such a system slowly evolves among branch states,
jumping over microscopic energy barriers due to the influence 
of thermal noise. This phenomenon can be regarded as an example of  
slow dynamics driven by a perturbation
along a neutrally stable manifold 
consisting of an infinite number of branch states.
\end{abstract}


\maketitle

An Ising model with a ferromagnetic interaction has a simple phase
space, and below some critical temperature 
it only possesses two macroscopic states, two oppositely 
directed ferromagnetic states which have the same energy.
Contrastingly, in the case of ferromagnetic coupled oscillators with
coupling functions containing strong higher harmonics, 
each oscillator is multi-stable, and as a result, 
the configuration observed in a simulation 
will consist of individual clusters characterized by different phases. 
Daido showed through numerical simulations 
that there exist many stable states, whose number scales as
 $\sim$ $O(\exp N)$ (where $N$ is the system size)\cite{daido2}. 
This complex mutual entrainment is called 
{\it multibranch entrainment} (MBE).
This phenomenon share the same mechanism of so-called {\it clustering},
which has been found in various systems, e.g., 
globally coupled maps \cite{Kaneko1} 
and pulse-coupled oscillators \cite{Ernst}.
When such systems exhibit periodic behavior,
their dynamics can be described by phase equations 
with coupling functions containing strong higher harmonics \cite{Ernst}.

It is a general principle of statistical mechanics that 
ergodicity can be broken only by {\it macroscopic} barriers 
of the free energy whose heights scale as $\sim$
$O(N)$. Here, we evaluate the free energy of a system 
of phase oscillators with coupling functions containing 
strong higher harmonics. Then, we demonstrate that 
MBE means ergodicity breaking caused by 
{\it microscopic} free energy barriers whose heights scale as $\sim$ $O(1)$.
As this result reveals, MBE seems to be in conflict with a basic principle 
of statistical physics, because a system can jump over 
these microscopic barriers due to the influence 
of thermal noise. 
Next, we apply the {\it moment approach} \cite{perez,bonilla}
to such a system and study properties of its relaxation process.
Then we elucidate the effect of microscopic energy barriers and 
demonstrate thermodynamical instability of MBE.

The ferromagnetic Ising model and the ferromagnetic analog network,  
governed by a Markovian process and a continuous process,  
are in the general class of time-dependent Ginzburg-Landau (TDGL) models \cite{Hohenberg}, 
because in such systems, the dynamics of macroscopic order parameters are described 
by the steepest-descent of the free energy.
A ferromagnetic coupled oscillator, however, is not in the class of TDGL models.
In this paper, we construct a Markovian system that has 
the same free energy as a ferromagnetic coupled oscillator but, nonetheless,  belongs to the class of TDGL models. Then, comparing this TDGL system
and non-TDGL ferromagnetic coupled oscillator system, 
we make clear that microscopic energy barriers 
have no effect on the relaxation of the system in TDGL models we consider but that 
they cause the relaxation to become slow in non-TDGL models we consider.
Our results suggest that such slowing of relaxation is characteristic
only of non-TDGL models.

We propose the following phase model with uniform all-to-all coupling by 
modifying Daido's model \cite{daido2}:  
\begin{eqnarray}
\frac{d \phi_i}{d t} &=& \omega_i - \frac{1}{N} \sum_{j=1}^{N} 
[ (1-a) \sin\left( \phi_i - \phi_j\right) \nonumber \\ 
& &\hspace{0.5cm}+ a \sin\left( 2 (\phi_i - \phi_j)\right)] + n_i(t). \label{Eq.sys1}
\end{eqnarray}
Here $\phi_i$ is the phase of the $i$th oscillator 
(with a total of $N$), and $\omega_i$ represents its quenched 
natural frequency.
The natural frequencies are randomly distributed 
with a density represented by $g(\omega)=(2\pi\sigma^2)^{-1/2} \exp(-\omega^2/2\sigma^2)$.
The parameter $a$ denotes the degree of mixture of the fundamental harmonic component
and the higher harmonic component in the coupling functions
($0 \leq a \leq 1$).
This model belongs to the class of ferromagnetic systems 
and does not exhibit frustration in the mutual interaction.
The function $n_i$ represents a Gaussian independent white-noise process 
satisfying 
$\left<n_i(t) \right> = 0, \ \ 
\left<n_i(t)n_j(t') \right> = 2 T \delta_{ij} \delta(t-t')$, 
where $T$ is the temperature of the system, and the inverse temperature
is defined as $\beta=1/T$.
Daido used a coupling of the form $\sin x + a \cos 2 x$
instead of that in Eq. (\ref{Eq.sys1}). In the case of Daido's model, when 
 $a$ gradually increase, another stable solution suddenly 
appears apart form original stable solution in coupling function. 
This sudden appearance of another stable solution also occurs in our model. 
Thus, the bifurcation structure of 
Daido's model is qualitatively equal to that of our model.
Moreover, when there is no frequency disorder, detailed balance holds in
our model. Therefore, we can analyze the MBE of our model 
in a transparent manner. 

For convenience, we define $s_i = \exp(i \phi_i)$ and 
rewrite Eq. (\ref{Eq.sys1}) in  the following form 
with bare mean fields:
\begin{eqnarray}
\frac{d \phi_i}{dt} &=&  \omega_i + \frac{1}{2i} [(1-a)\left(\overline{s}_i m_1 - s_i \overline{m_1}\right) \nonumber \\
& &\hspace{0.5cm}+ a \left(\overline{s}^2_i m_2 - s^2_i \overline{m_2}\right)] 
+ n_i(t). \label{Eq.sys2}
\end{eqnarray}
Here the overline denotes the complex conjugate.
The quantity $m_1$ is a mean field consisting of the fundamental harmonics  
$\{s_j\}_{j=1,\cdots, N}$, and $m_2$ is a mean field 
consisting of the higher harmonics $\{s_j^2\}_{j=1,\cdots, N}$:
$m_1 = \frac{1}{N}\sum_j^{N} s_j$;  
$m_2 = \frac{1}{N}\sum_j^{N} s^2_j$.

Equation (\ref{Eq.sys2}) obeys 
the Fokker-Planck equation
\begin{eqnarray}
& &\frac{\partial p(\phi|\omega, t)}{\partial t} =
T \frac{\partial^2 p(\phi|\omega, t)}{\partial \phi^2}  
- \frac{\partial}{\partial \phi} \left(v(\phi, \omega) p(\phi|\omega, t) \right), \label{Fokker-Planck} \\
& &v(\phi, \omega) = \omega \nonumber\\
& &\ \ + \frac{1}{2i} \left((1-a)\left(\overline{s} m_1 - s \overline{m_1}\right) + a \left(\overline{s}^2 m_2 - s^2 \overline{m_2}\right)\right), \label{flow} 
\end{eqnarray}
where $p(\phi|\omega, t)$ denotes the one-oscillator probability density 
at time $t$. The site index $i$ of $\omega$ and $\phi$ can 
be omitted, because Eq. (\ref{Eq.sys2}) has already been  reduced to 
a one-body system.  

We derive the macroscopic dynamics of the system 
by using the {\it moment approach} \cite{perez,bonilla}.
We define the moments of $p(\phi|\omega, t)$ as
\begin{eqnarray}
\lambda_n(\omega, t) = \int_{0}^{2\pi} d\phi p(\phi|\omega, t) \exp(-i n \phi).
\end{eqnarray}
The motion of these moments obeys the simple equation
\begin{eqnarray}
& &\frac{d \lambda_n(\omega,t)}{dt} = -T n^2 \lambda_n(\omega, t) 
+ i n \omega \lambda_n(\omega, t)\nonumber \\
& &-\frac{n(1-a)}{2}\left(m_1 \lambda_{n+1}(\omega, t)
-\overline{m_1} \lambda_{n-1}(\omega, t)\right)
\nonumber \\
& &- \frac{an}{2}\left(m_2 \lambda_{n+2}(\omega, t)
-\overline{m_2} \lambda_{n-2}(\omega, t)\right), \label{Eq.md1}
\end{eqnarray}
which is derived through a moment expansion of Eq. (\ref{Fokker-Planck}).
Equation (\ref{Eq.md1}) possesses a hierarchal structure, because 
the $n$th moment depends on neighboring moments 
[the $(n-2)$th, $(n-1)$th, $n$th, $(n+1)$th and $(n+2)$th moments].
The $0$th moment is found trivially to be $\lambda_0(\omega, t) = 1$.
The mean fields $m_1$ and $m_2$ are given by  
$\overline{m_1} = \int d\omega g(\omega) \lambda_{1}\left(\omega, t \right), 
\ \ 
\overline{m_2} = \int d\omega g(\omega) \lambda_{2}\left(\omega, t \right)$.
When there is no frequency disorder, we obtain
$\overline{m_1} = \lambda_{1}\left(0, t \right)$ and 
$\overline{m_2} = \lambda_{2}\left(0, t \right)$.
If $T \neq 0$, higher moments rapidly decay,
since the coefficient of the damping factor in Eq. (\ref{Eq.md1})
is $T n^2$. Therefore, even if higher moments are omitted  
in the numerical calculation of Eq. (\ref{Eq.md1}), 
highly accurate results can be obtained. 
For the convenience of later discussions,
we call the system without the thermal noise term the ``Liouville part'', 
because Eq. (\ref{Eq.md1}) without the diffusion part 
(i.e., the damping factor) is equivalent to the so-called Liouville equation.
The diffusion part in Eq. (\ref{Eq.md1}) can be regarded as a perturbation
of the Liouville equation.

When there is no frequency disorder [i.e. when  
$g(\omega)= \delta(\omega)$], detailed balance holds,
since the coupling function is an odd function \cite{aonishi,bonilla}.
Thus, in the limit $\sigma \rightarrow 0$,
this model can be mapped to the equilibrium system.
In this case, there exists a free energy, which enables us to 
apply equilibrium statistical mechanical methods 
to the system. Then we can obtain the Gibbs distribution and 
partition function
\begin{eqnarray}
P(\mbox{\boldmath$\phi$}) 
= e^{ - \beta H(\mbox{\boldmath$\phi$})}/ Z, \ \ 
Z = \int_{0}^{2\pi} d\mbox{\boldmath$\phi$} e^{ - \beta H(\mbox{\boldmath$\phi$})}, \label{Eq.dis}
\end{eqnarray}
in equilibrium \cite{bonilla}, 
where $H$ denotes the Hamiltonian expressed by 
\begin{eqnarray} 
& &H(\mbox{\boldmath$\phi$}) = \nonumber \\
& &-\sum_i \left( 
\frac{1-a}{2}\left( \overline{s}_i m_1 + s_i \overline{m_1} \right)
+\frac{a}{4}\left( \overline{s}^2_i m_2 +  s^2_i \overline{m_2}\right)
\right).
\end{eqnarray}
We thus obtain the free energy  
$f =  - 1/(\beta N) \log Z$. 
Using the saddle-point method to evaluate the integral in $Z$,
for large $N$, we derive the free energy as  
\begin{eqnarray} 
& &f(m_1,m_2,\overline{m_1},\overline{m_2})
= \frac{1-a}{2} |m_1|^2 +\frac{a}{4} |m_2|^2
\nonumber \\
& &\hspace{0.4cm}-\frac{1}{\beta}  \log\left<e^{\beta \left(\frac{1-a}{2}
\left(\overline{s} m_1 + s  \overline{m_1} \right) +\frac{a}{4} \left(\overline{s}^2  m_2  + s^2  \overline{m_2} \right) \right)}\right>_{\phi}. 
\label{Eq.free_sad}  
\end{eqnarray}
In the limit $T\rightarrow 0$, this becomes 
\begin{eqnarray} 
f(m_1,m_2,\overline{m_1},\overline{m_2})= \frac{1-a}{2} |m_1|^2 + \frac{a}{4} |m_2|^2 = \frac{H}{N}. \label{Eq.free_sad2} 
\end{eqnarray}
This is the Hamiltonian normalized with respect to $N$.

\begin{figure}[ht]
\includegraphics[height=2.4cm]{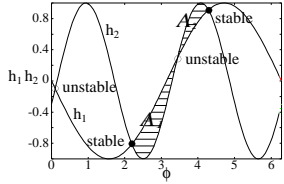}
\caption{Sketch of microscopic bistability.}
\label{sc}
\end{figure}

Let us consider the fixed points of Eq. (\ref{Eq.sys2}) in the limit  
$T \rightarrow 0$ and $\sigma \rightarrow 0$.  
Figure \ref{sc} displays a typical example of 
a fundamental harmonic component
and a higher harmonic component of Eq. (\ref{Eq.sys2}) 
in the case $\frac{d\phi_i}{dt}=0$. 
Namely, figure \ref{sc} shows two curves  
$h_1(\phi_i) = -\frac{1-a}{2i} \left(\overline{s}_i m_1 - s_i
\overline{m_1}\right)$ and 
$h_2(\phi_i) = \frac{a}{2i} \left(\overline{s}^2_i m_2 - s^2_i \overline{m_2}\right)$.
If $a$ is sufficiently large, Eq. (\ref{Eq.sys2})
admits four solutions corresponding to intersection points 
of the fundamental harmonic component and the higher harmonic component,
as shown in Fig. \ref{sc}. The filled circles and open circles correspond 
to stable solutions and unstable solutions, respectively.
We thus find that this system is microscopically bistable. 
The two enclosed areas $A_1$ and $A_2$ in Fig. \ref{sc} 
are, respectively, equal to the energies of the two stable solutions,
as measured with respect to the unstable solution between them.
The microscopic energy barrier discussed in introduction 
is the energy barrier between these stable solutions.
It is a principle of statistical physics that 
in a situation like this, only the 
stable solution with the larger enclosed area can be realized,
because only the stable solution with the lowest  
energy is thermodynamically stable. 
This rule is referred to as {\it Maxwell's rule}
\cite{fukai4}. Note that two stable solutions have the
same energy only if $a=1$.
However, if system is non-ergodic, multiple states can coexist, 
and Maxwell's rule does not hold. 
In this case, depending on the initial conditions, each oscillator 
comes to belong to one of two clusters. 
Then the total number of stable solutions
in the whole system is $O(2^N) \sim O(\exp N )$. 
This is the situation involving MBE.

\begin{figure}[ht]
\includegraphics[width=3.5cm]{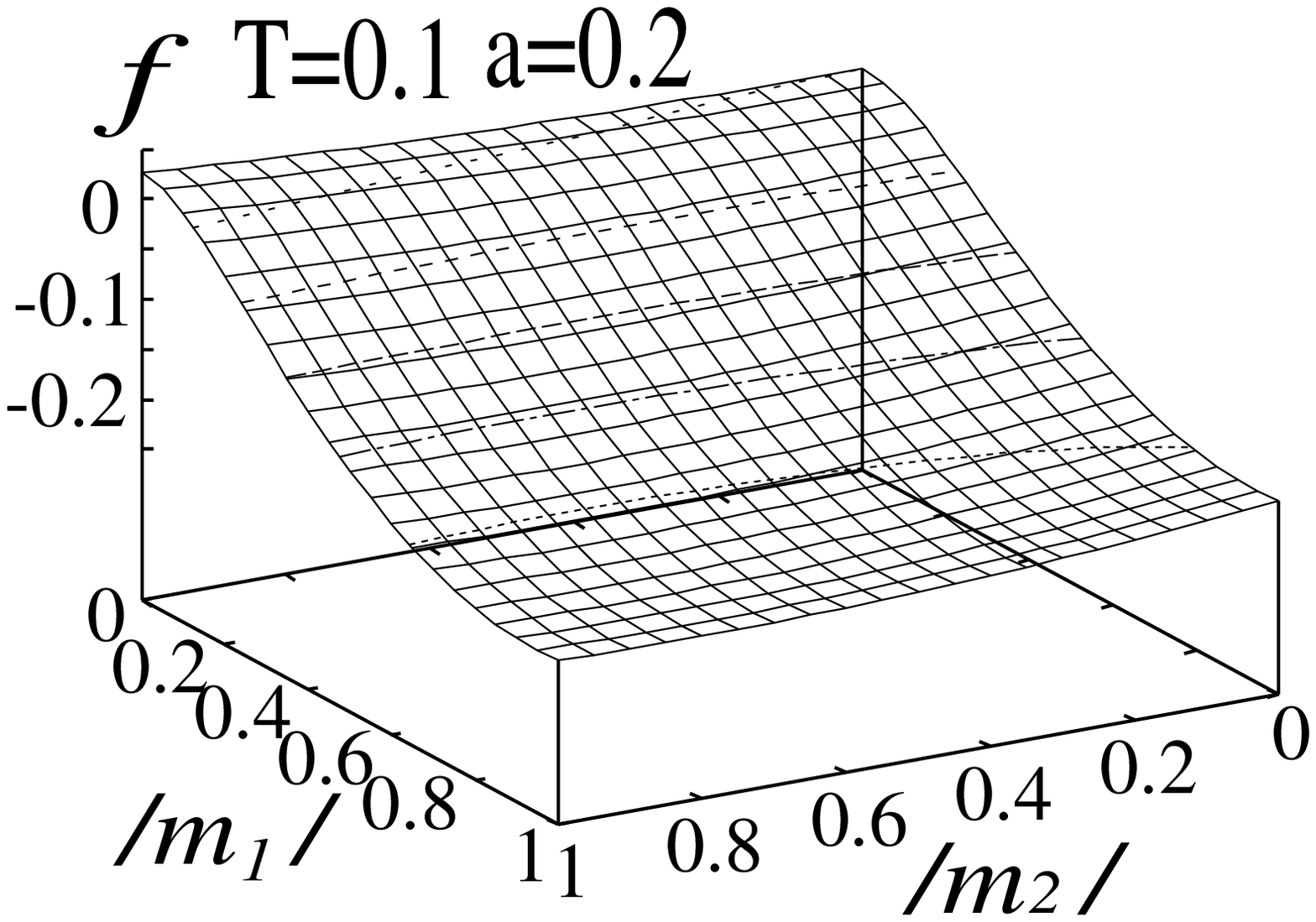}
\includegraphics[width=3.5cm]{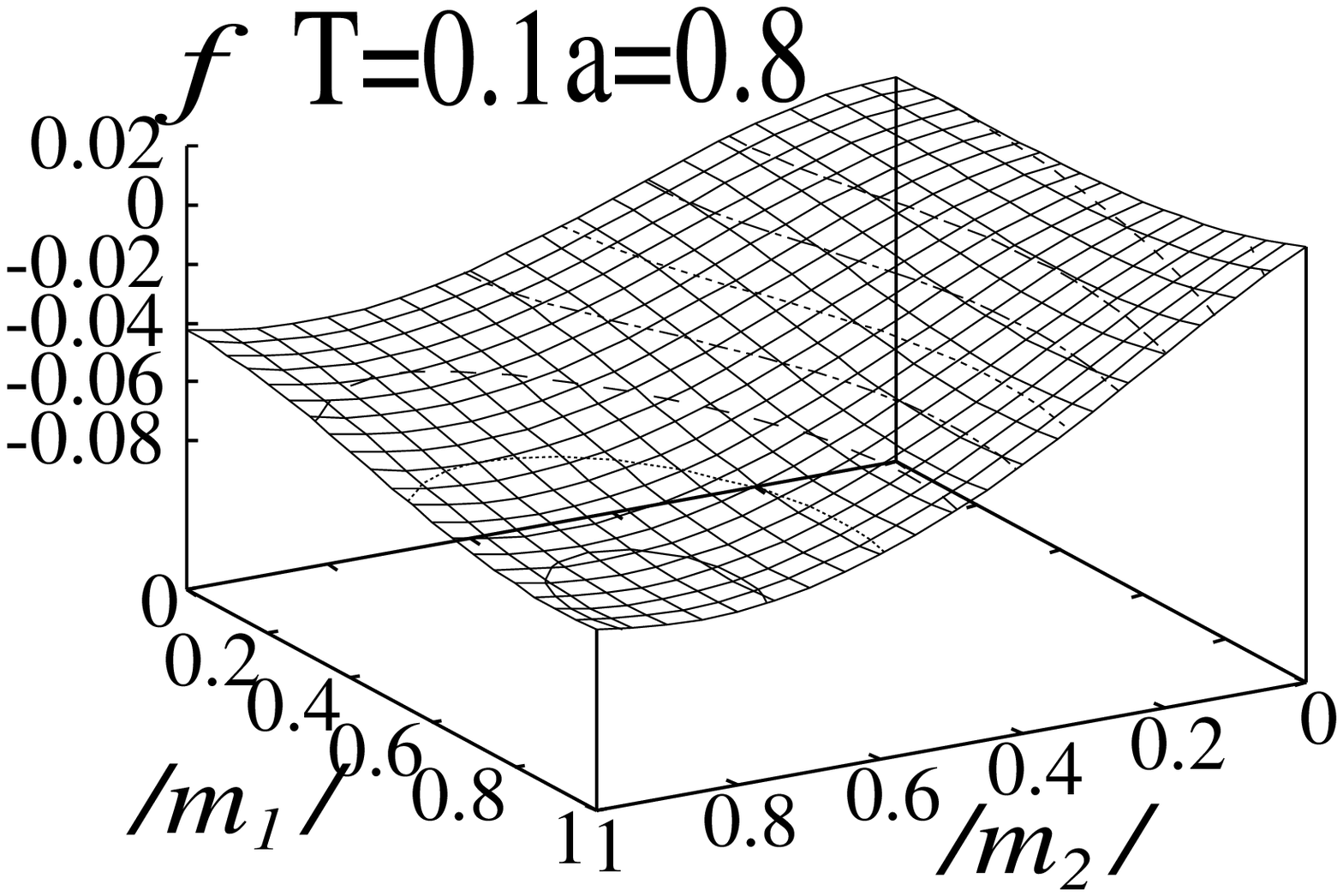} 
\caption{Macroscopic landscape of the free energy.}
\label{cd_free}
\end{figure}

Here, we demonstrate that microscopic multi-valley of the free energy 
is not reflected in the macroscopic structure of the free energy.
Figure \ref{cd_free} displays the macroscopic landscape of the free energy
(Eq. (\ref{Eq.free_sad})) as a function of $|m_1|$ and $|m_2|$.
The left figure corresponds to the case of a monostable system ($a=0.2$),
and the right figure corresponds to the case of a bistable system 
($a=0.8$). There are no any local minima of the two energy surfaces. 
Thus, it is seen even in the case of a bistable system, the free energy 
does not possess a multi-valley structure macroscopically.
From Eq. (\ref{Eq.free_sad2}), it is clear that, even in the limit 
$T\rightarrow 0$,  
there is no macroscopic multi-valley structure in the free energy.
Note that in the limit $T \rightarrow 0$, the critical value of 
$a$, representing the boundary between monostability and bistability,
is given by $a_c = 0.467$. 

As the macroscopic structure of the free energy 
reveals, it seems that the ergodicity breaking
by microscopic barriers does not occur. 
However, when $g(\omega) = \delta(\omega)$, $T\rightarrow 0$ and $a>a_c$,
$\lambda_n(0, \infty) =  r + (1-r)(-1)^n$ corresponding to 
some branch state is a stable solution of Eq. (\ref{Eq.md1}).
In this case, the mean fields $m_1$ and $m_2$ are given by 
$m_1 = 2 r - 1$ and $m_2 = 1$.
These moments yield the probability density function 
$p(\phi|0,\infty) = r \delta(\phi)+(1-r)\delta(\phi-\pi)$,
where $r$ denotes the ratio of oscillators at $\phi = 0$
to those at $\phi = \pi$. This solution is satisfied 
for any $r$ $(0 \leq r \leq 1)$. 
In this case,  these solutions form a line attractor  
parameterized by $r$:
$M = \left\{ r + (1-r)(-1)^n | 0 \leq r \leq 1 \right\}$ .
Within this attractor, all of the solutions
are neutrally stable. We thus find that in the limit $N \rightarrow \infty$, 
the set of solutions forming MBE can be parameterized 
by the single variable $r$.

Therefore, in the limit $T\rightarrow 0$, the global 
ergodicity demonstrated by the macroscopic structure of the free energy 
seems to be in conflict with 
the existence of $M$ revealed by the moment approach.
In the following numerical calculations, we evaluate the dynamical 
properties of this system at low temperature, and then answer the question
what happens in the limit $T\rightarrow 0$.
In the numerical simulations we now discuss, the initial
conditions of the system were assigned as random numbers 
with the density function
$p(\phi|\omega,0) = q \delta(\phi)+(1-q)\delta(\phi-\pi)$,
where $q$ denotes the ratio of oscillators at $\phi = 0$
to those at $\phi = \pi$. Thus, initial conditions were chosen 
on the manifold $M$. 

\begin{figure}[ht]
(a)\hspace{-0.2cm}\includegraphics[width=3.8cm]{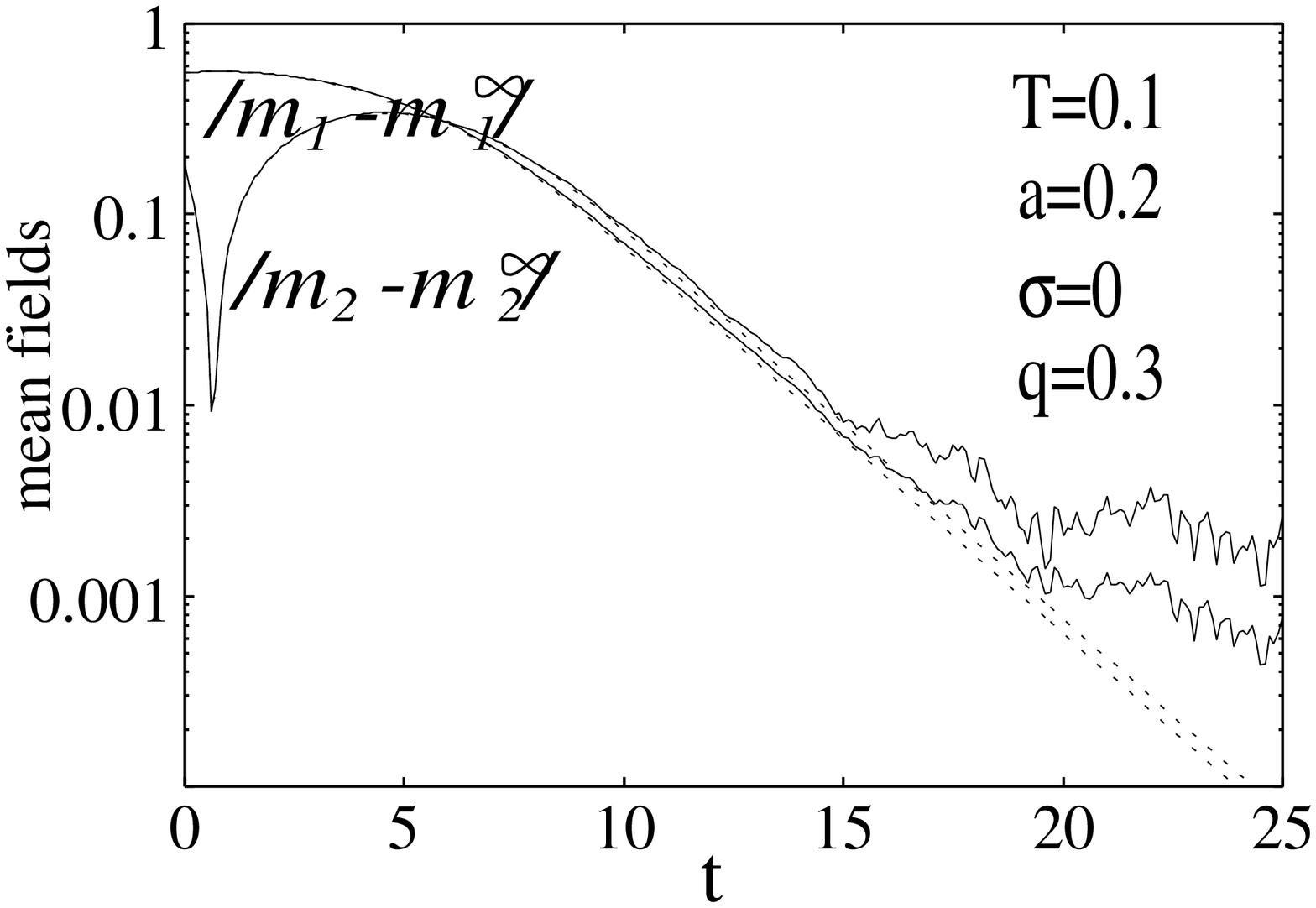}
\includegraphics[width=3.8cm]{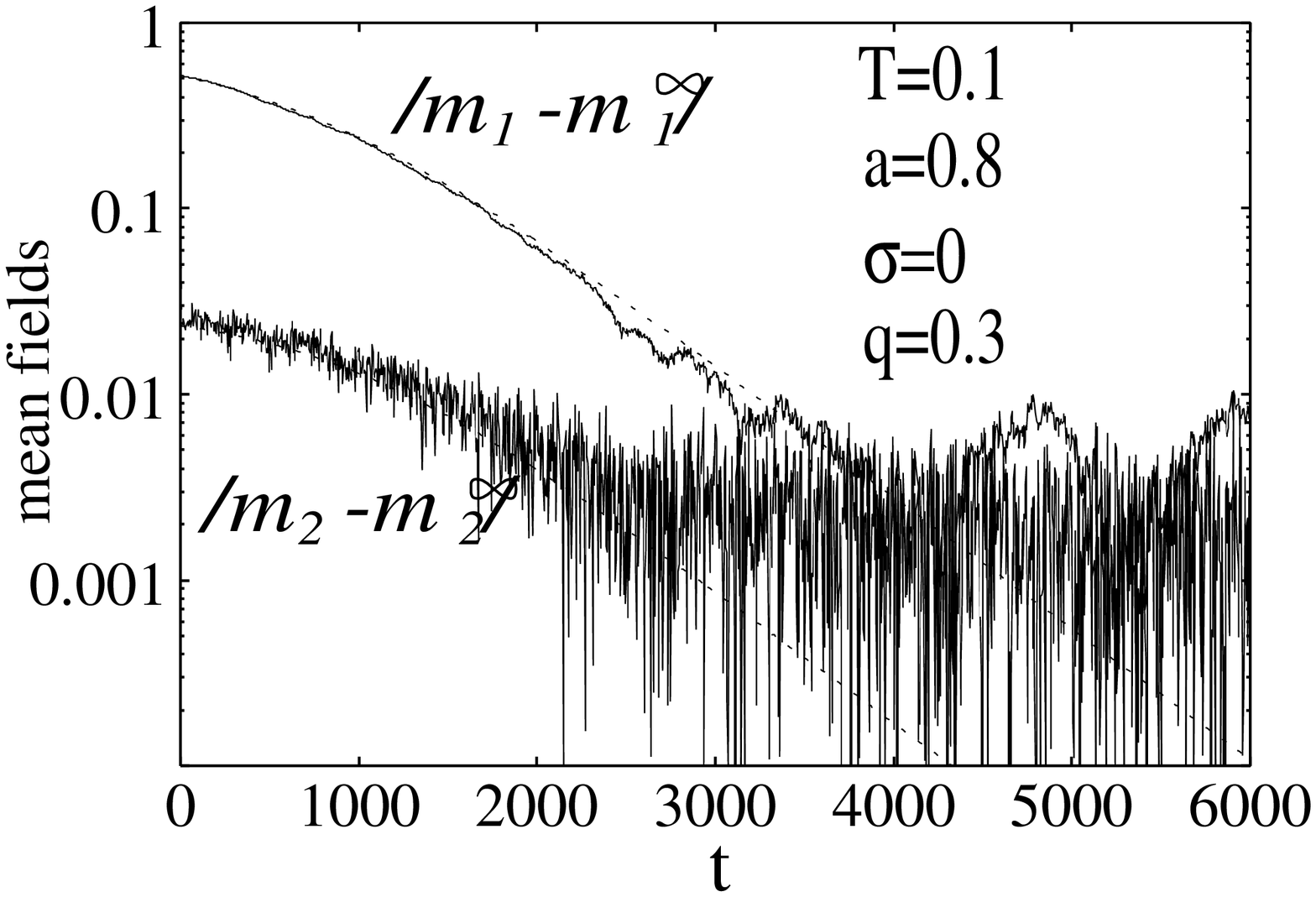}\\
(b)\hspace{-0.2cm}\includegraphics[width=4cm]{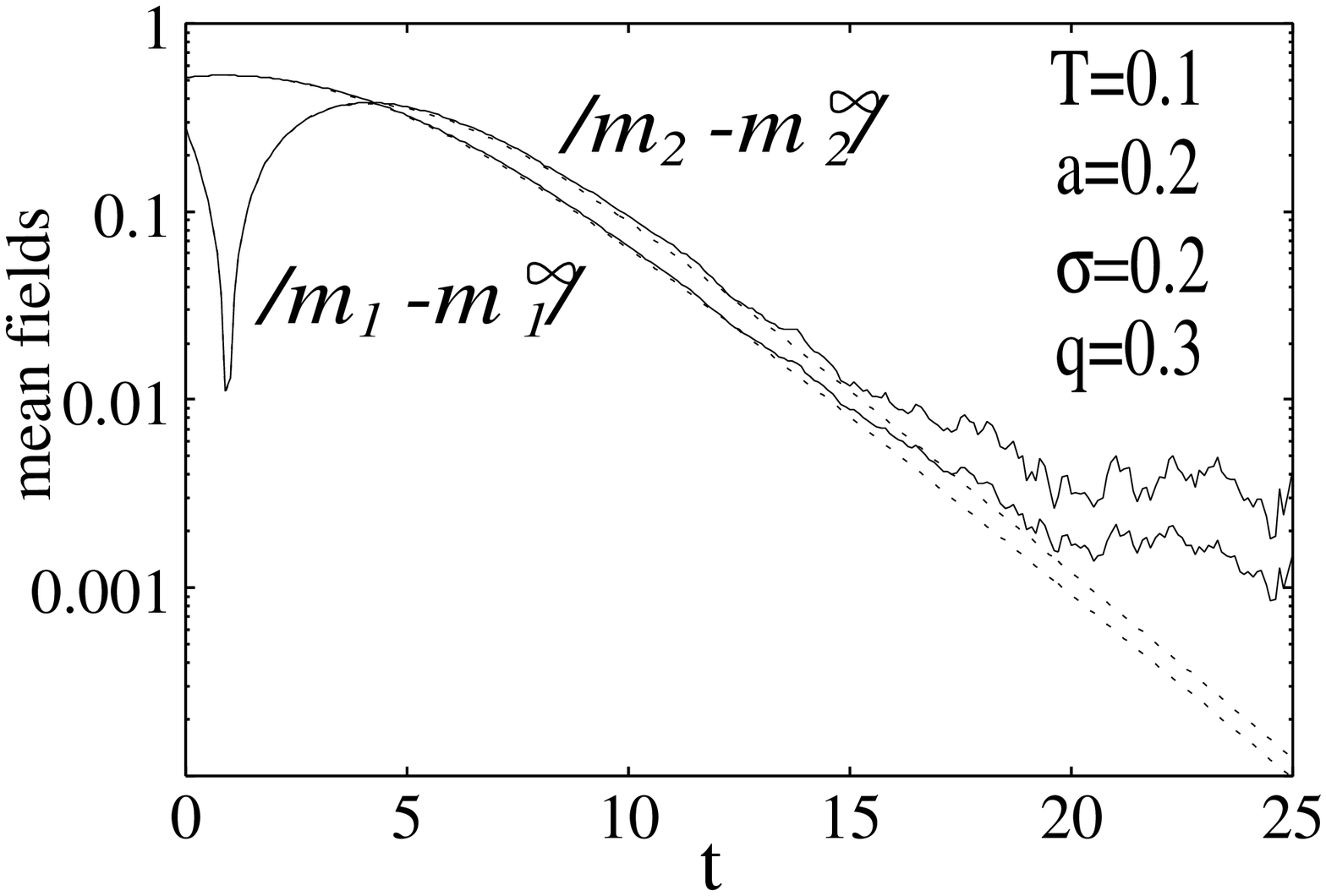}
\includegraphics[width=3.8cm]{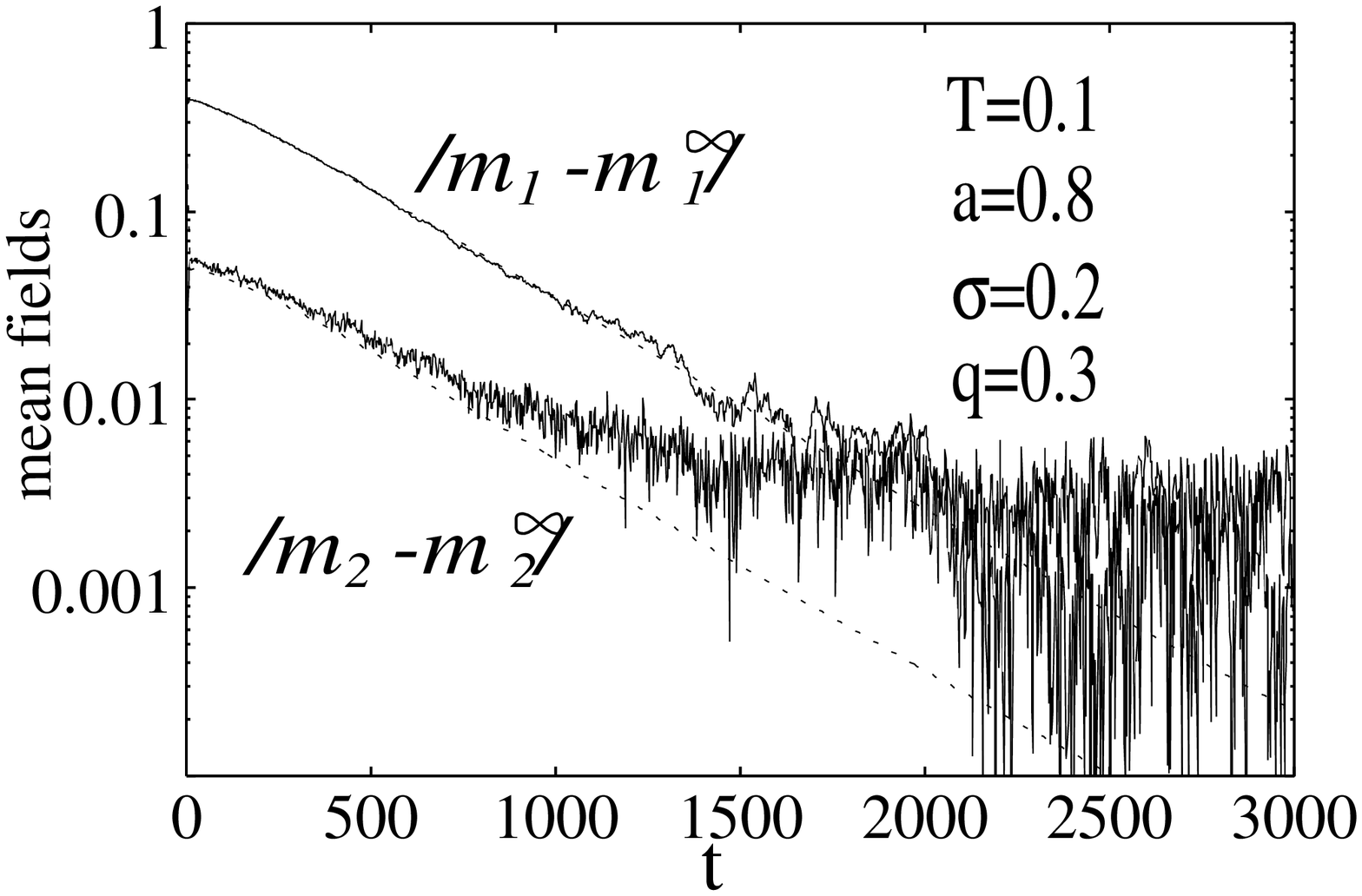}
\caption{Temporal behavior of $m_1-m_1^{\infty}$ and $m_2-m_2^{\infty}$:
(a) $\sigma = 0$. (b) $\sigma = 0.2$. 
The dashed curves were obtained analytically, while the 
solid curves were obtained from numerical simulations.}
\label{gin_dy}
\end{figure}

Figures \ref{gin_dy}(a)  displays  
the temporal behavior of $|m_1-m_1^{\infty}|$ and $|m_2-m_2^{\infty}|$,
where $m_1^{\infty}$ and $m_2^{\infty}$ denote values of $m_1$
and $m_2$ in an equilibrium state corresponding to the minimum point of
the free energy shown in Fig. \ref{cd_free}. 
The dashed curves in these figures are the analytical results 
obtained from Eq. (\ref{Eq.md1}), and the solid curves are numerical
results. As these figures reveal,
the analytical results fits in with the simulation results.

\begin{figure}[ht]
(a)\hspace{-0.2cm}\includegraphics[width=3.6cm]{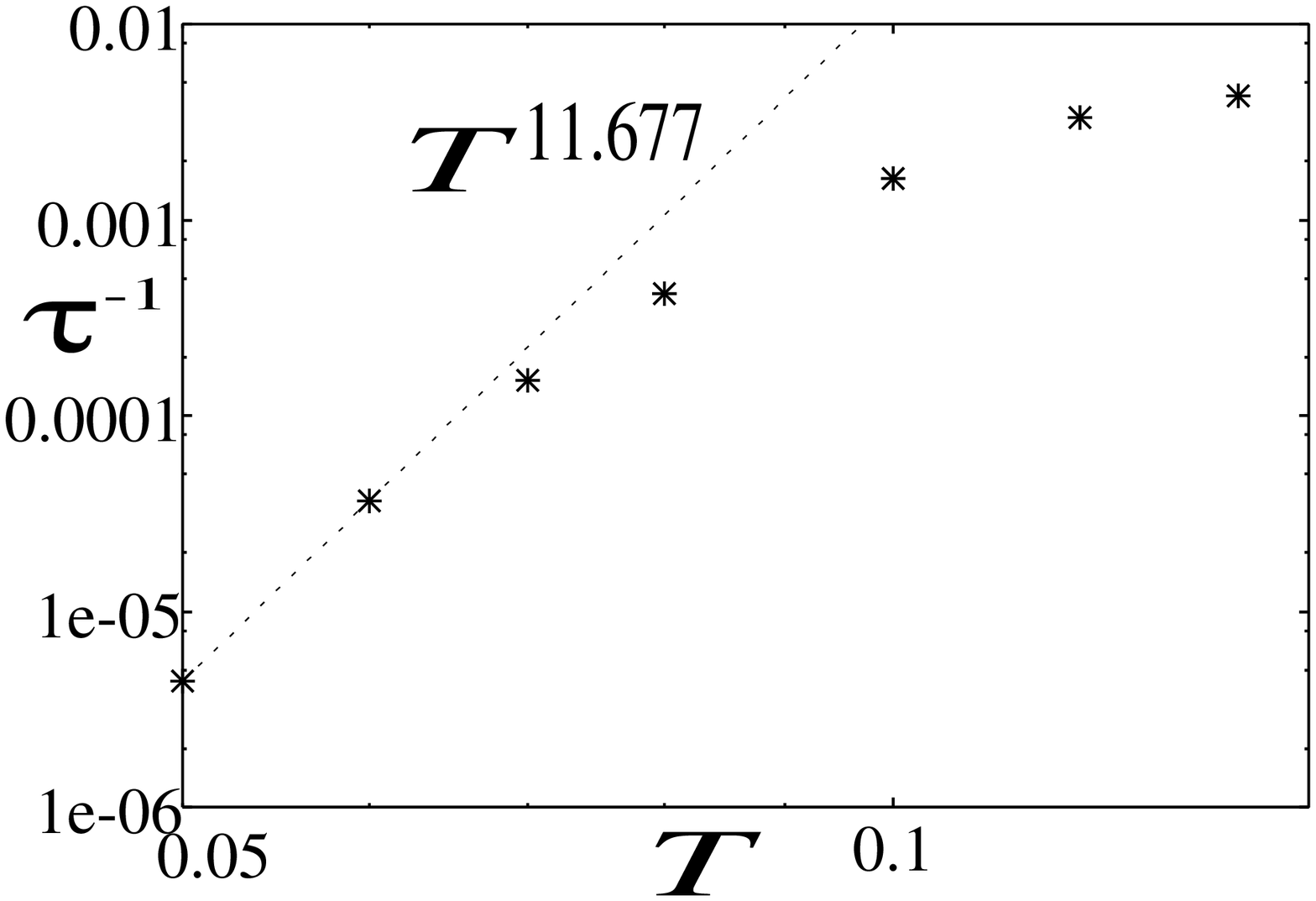}
(b)\hspace{-0.2cm}\includegraphics[width=3.6cm]{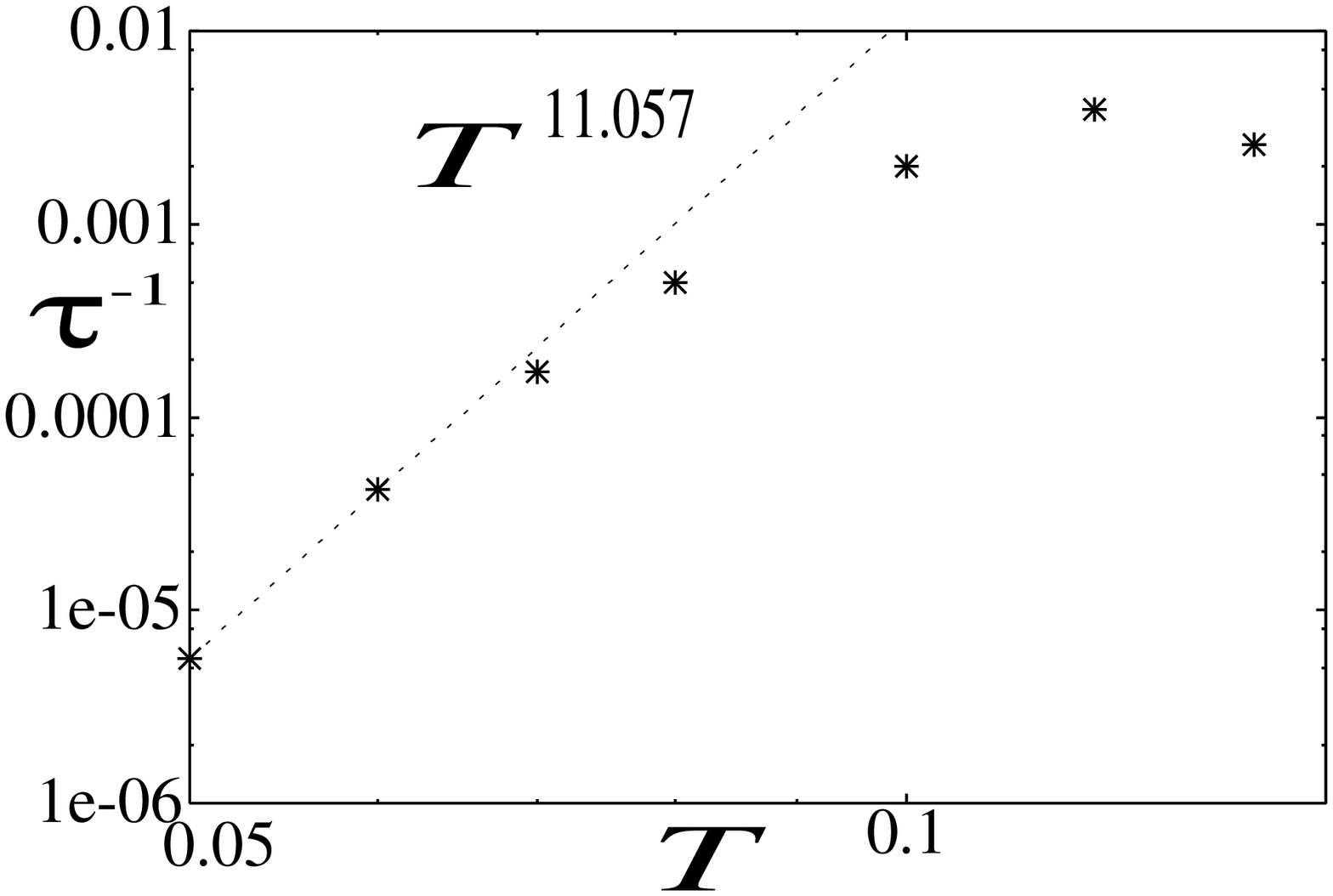}\\
(c)\hspace{-0.2cm}\includegraphics[width=3.6cm]{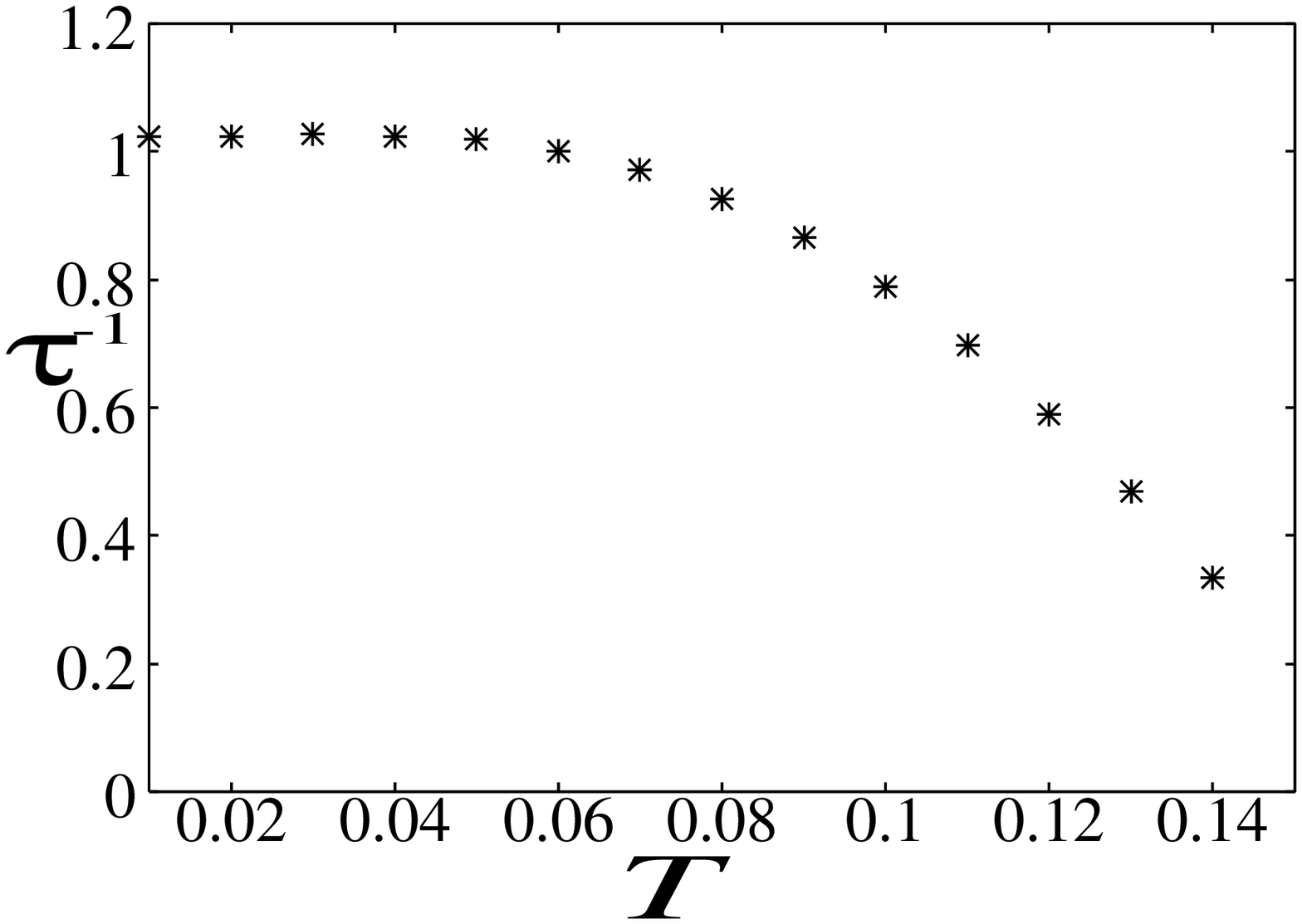}
\caption{Time constant $\tau$ as a function of $T$:
(a) $\sigma=0, a=0.8, q =0.3$;
(b) $\sigma=0.1, a=0.8, q =0.3$;
(c) Markovian system ($\sigma=0, a=0.8, q=0.3$).}
\label{gin_tau}
\end{figure}

We now calculate the time constant $\tau$ of the system  
using logarithm plots. According to Fig. \ref{gin_dy}(a), 
the relaxation process to equilibrium states is characterized by 
the exponential convergence. The time constant of this convergence,
$\tau$, for the bistable system ($a=0.8$) is about $100$ times larger than 
that for the monostable system ($a=0.2$). 
We also evaluate the time constant $\tau$ as a function of $T$ 
in the case of a bistable system ($a=0.8$). 
As Fig. \ref{gin_tau}(a) reveals, $\tau$ diverges to $\infty$ in the limit
$T\rightarrow 0$. 

\begin{figure}[ht]
\includegraphics[height=1.8cm]{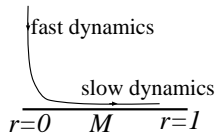}
\caption{Sketch of dynamical flow around $M$.}
\label{mani}
\end{figure}

Note that in the initial stages of the relaxation process,
when initial states are far from $M$, 
the speed of convergence for the system with $a=0.8$ is 
almost the same as that for the system with $a=0.2$.
During the approach to the equilibrium state, 
the relaxation process suddenly switches 
from a process of fast dynamics to a process of slow dynamics (see Fig. \ref{mani}).
The mechanism responsible for this switch is the following. 
Since the thermal perturbation (the diffusion part in Eq. (\ref{Eq.md1}) ) 
is very small, for solutions far from $M$,
the perturbation is negligible, and they are attracted rapidly to 
$M$, just as in the unperturbed system.
(This rapid attraction is called the ``fast dynamics''.) 
Now note that the Liouville part vanishes on $M$. For this reason, once
solutions enter the 
neighborhood of $M$, where the effects of the Liouville part
and the perturbation become comparable, 
the solution begins to move slowly along $M$ toward the ground state.
(This is called the ``slow dynamics''.) 


In order to compare the non-TDGL system considered above to
a TDGL system, we evaluated 
the time constant $\tau$ for a Glauber system.  
This system was constructed to have 
the same Gibbs distribution as Eq. (\ref{Eq.dis})
in the equilibrium state. 
Thus, its macroscopic dynamics are described 
by the steepest-descent of the free energy given in Eq. (\ref{Eq.free_sad}).
As Fig. \ref{gin_tau}(c) reveals, in this case $\tau$ converges to some finite value 
as $T\rightarrow 0$, even for a bistable system ($a=0.8$).
In the limit $T \rightarrow 0$, 
the Glauber system chooses only the stable solution 
with the lower energy. We thus find that microscopic energy barriers do not affect 
the relaxation process of the Glauber system.
This contrasts with what we have found for the relaxation process of
the continuous system Eq. (\ref{Eq.sys1}), which is hampered 
by the microscopic energy barriers, resulting in a slow
(infinitely slow in the $T\rightarrow 0$ limit) relaxation 
to the ground state.

In order to study the effect of the frequency disorder,
we also measured the temporal behavior of $|m_1-m_1^{\infty}|$ and
$|m_2-m_2^{\infty}|$ in the case $\sigma \neq 0$ (see Fig. \ref{gin_dy}(b).)
Here, $m_1^{\infty}$ and $m_2^{\infty}$ denote values of $m_1$
and $m_2$ in a macroscopic steady state \cite{daido2}.
Even if there is frequency disorder, the time constant of this convergence,
$\tau$, for the bistable system ($a=0.8$) is about $100$ times larger than 
that for the monostable system ($a=0.2$). 
As shown in Fig. \ref{gin_tau}(b), $\tau$ diverges to 
$\infty$ in the limit $T\rightarrow 0$. 

In conclusion, the system we have considered evolves slowly 
through a series of branch states
to the ground state, jumping over 
microscopic energy barriers through the influence of thermal noise. 
The time constant $\tau$ characterizing this relaxation process  
diverges to $\infty$ in the limit $T\rightarrow 0$.
From the macroscopic viewpoint, this phenomenon can be regarded 
as slow dynamics driven by a weak thermal perturbation
along the neutrally stabile manifold $M$ consisting of 
an infinite number of branch states.
From these results, we can conclude that, in the long term,
MBE is unstable thermodynamically, but 
in the short term, the system preserves phase patterns highly correlated
with the initial conditions, because its relaxation time is very long.

There exist some theoretical studies of coupled oscillators 
based on the assumption that    
the effect of the frequency disorder is equivalent to 
that of thermal noise \cite{Vicente,Park}.
According to the present results, we can conclude that 
the effect of quenched frequency disorder is far different from that of
thermal noise \cite{Braiman,Hentschel}. 

\bibliography{ref.bib}

\end{document}